\documentclass[12pt]{article}
\usepackage[english]{babel}

\usepackage[colorlinks=true,backref=true,linkcolor=blue,linktocpage=true,anchorcolor=black,citecolor=blue,filecolor=black,menucolor=black,pagecolor=black,urlcolor=blue]{hyperref}

\usepackage{amsmath}
\usepackage{bm}
\usepackage{amsfonts}
\usepackage{graphicx}
\usepackage{caption}
\usepackage{subcaption}
\usepackage{comment}
\usepackage{amssymb}
\usepackage{cite}
\usepackage{authblk}
\usepackage[usenames,dvipsnames]{xcolor}
\usepackage{tikz}
\usetikzlibrary{patterns}
\usepackage{mathtext}
\usepackage{pgfplots}
\usetikzlibrary{pgfplots.groupplots}
\usepackage{tkz-euclide}
\usetkzobj{all}

\usepackage{rotating}
\usepackage{accents}

\newcommand{\nn}{\nonumber}

\usepackage{authblk}
\title{
BKL oscillations in 2+1 space-time dimensions
}
\author[1]{Philipp Fleig}
\author[2,3]{Vladimir A. Belinski}
\affil[1]{Max Planck Institute for Dynamics and Self-Organization, 37077 G\"ottingen, Germany}
\affil[2]{International Center for Relativistic Astrophysics Network, 65122 Pescara, Italy}
\affil[3]{Institut des Hautes \'Etudes Scientifiques, 91440 Bures-sur-Yvette, France}

\begin{document}

\maketitle
\abstract{
We investigate the question whether there are cosmological models in $2+1$ space-time dimensions which exhibit dynamics similar to BKL oscillations, as the cosmological singularity is approached. Based on intuition, we conceive a toy model which displays such oscillatory dynamics. We show that in the phase space of this model, the cosmological singularity is represented by a separatrix curve and discuss the model's dynamics within the cosmological billiards picture. 
Finally, we offer a physical interpretation for a family of similar cosmological models in terms of the topological degrees of freedom of gravity in $2+1$ dimensions.
}

\section{Introduction}
Finding a description of the dynamics of space-time in the vicinity of a space-like singularity may be considered as one of the most premier problems of theoretical physics. While a complete solution of the problem may quite possibly involve a quantised form of the gravitational field, a few important results have already been obtained within classical General Relativity theory. One of these is the Belinski-Khalatnikov-Lifshitz (BKL) conjecture put forward in~\cite{Lifshitz:1963ps,Belinsky:1970ew,Belinsky:1982pk,Belinsky:1988mc}. In a nutshell, according to BKL, space-time in the vicinity of a cosmological singularity undergoes an infinite sequence of stochastic oscillations. Since their original proposal, the work of BKL has been extended further greatly and the stochastic nature of space-time near a cosmological singularity has been identified as a generic phenomenon, present also in theories of gravity other than General Relativity. While the BKL conjecture has been studied for a wide range of extensions of General Relativity (and possibly most notably String Theories, see e.g.~\cite{DH:chaos_superstring,DH:oscillatory_behaviour}), the question whether oscillatory dynamics may also be found for gravity in $2+1$ dimensions, has to the best of our knowledge not been investigated so far. In fact, at first sight it might seem that the answer to this question has to be negative for the following reason: gravity in $2+1$ dimensions does not carry any local degrees of freedom, in other words, there are no gravitational waves, supposedly rendering the theory trivial. However it is also known, that in the case of non-trivial space-time topologies, gravity in $2+1$ dimensions can possess global (topological) degrees of freedom which are dynamical. For a detailed introduction see for example~\cite{Carlip:1995zj}. This insight opens up the possibility that in terms of global degrees of freedom it might be possible to construct models with a dynamics reminiscent of BKL and we will investigate this question here.      

The first goal of this work is to investigate to what extent the dynamics and technical notions found within the context of established models with BKL oscillations have an analogue in $2+1$ dimensions. 
We therefore begin in section~\ref{sec:BIXreview} with a brief review of the homogeneous, diagonal Bianchi IX model, introduce its Lagrangian form and discuss the occurence of BKL oscillations in the limit towards the singularity. Readers well familiar with the concepts of a Kasner solution, minisuperspace and cosmological billiards may wish to skip this section.
In section~\ref{sec:toymodel} we introduce a toy model in $2+1$ dimensions, and analyse the phase space dynamics of this model. Furthermore, we discuss its dynamics as the motion of a fictitious particle in minisuperspace. The second goal of the article is to make a proposal for how to interpret such cosmological models in $2+1$ dimensions physically. This is done in section~\ref{sec:physical_interpretation}.

\section{BKL oscillations in the BIX model}\label{sec:BIXreview}
In order to set the stage and introduce some notation we start with a concise review of BKL oscillations in the homogeneous, diagonal Bianchi IX model. For a more detailed review of the Bianchi IX model we refer the reader to~\cite{Belinski:reviewBKL} and to~\cite{DHN:reviewBKL} for an introduction to cosmological billiards.

The metric of 3+1-decomposed space-time takes the form
\begin{align}
ds^2=-(Ndx^0)^2 + g_{ij}dx^idx^j\,.
\end{align}
The homogeneous, diagonal Bianchi IX model is then specified by taking the spatial metric $g_{ij}$ to be of the form
\begin{align}
g_{ij}=h_{ab}\,\mathrm{e}_i^{\phantom i a}\mathrm{e}_j^{\phantom j b}\,,
\end{align}
where both the spatial indices $i,j$ and the frame indices $a,b$ take the values $1,2$ and $3$. The metric $h_{ab}$ is diagonal with its diagonal components parameterised by
\begin{align}\label{scalefacts}
a^2=e^{-2\beta^1}\,,\quad b^2=e^{-2\beta^2}\,,\quad c^2=e^{-2\beta^3}\,.
\end{align}
The variables $\beta^a$ in the exponents of the scale factors only depend on time $x^0$. For the Bianchi IX model the Dreibein, $\mathrm e_i^{\phantom i a}$, satisfies the standard structure equations\footnote{See for example~\cite{LL:classicalfields}:
\begin{align}
C^a_{\phantom abc}=e_{bcd}\,C^{da}=(\mathrm e^{\phantom \alpha a}_{\alpha,\beta}-\mathrm e^{\phantom \beta a}_{\beta,\alpha})(\mathrm e^{-1})_b^{\phantom b\alpha}(\mathrm e^{-1})_c^{\phantom c\beta}\,,
\end{align}
where $e_{abc}$ is the unit antisymmetric symbol with $e_{123}=1$. The Bianchi IX model is then obtained by setting the structure constants to $C^{11}=C^{22}=C^{33}=1$ with all other components vanishing.}. The time $x^0$ is related to the proper time $t$ via
\begin{align}
dx^0=-\frac{dt}{N}\,.
\end{align}
We fix the lapse function to $N=\sqrt{g}=abc$ and for this particular choice denote time by $\tau$ instead of $x^0$, so that
\begin{align}
d\tau=-\frac{dt}{\sqrt{g}}\,.
\end{align}
The singularity lies on a constant time slice, at proper time $t=0$.

With these definitions in place, we can write out explicitly the Lagrangian of the homogeneous, diagonal Bianchi IX model:
\begin{align}
L=T-V\,,
\end{align}
where the kinetic term takes the form
\begin{align}\label{Lagr_Mixmaster}
T&=\sum_i\left(\dot\beta^i\right)^2-\left(\sum_i \dot\beta^i\right)^2=-2\left(\dot\beta^1\dot\beta^2+\dot\beta^1\dot\beta^3+\dot\beta^2\dot\beta^3\right)
\end{align}
and the potential term is given by
\begin{align}\label{eq:potential}
V&=\frac12\left[e^{-4\beta^1}+e^{-4\beta^2}+e^{-4\beta^3}-2\left(e^{-2(\beta^1+\beta^2)}+e^{-2(\beta^1+\beta^3)}+e^{-2(\beta^2+\beta^3)}\right)\right]\,.
\end{align}
A dot denotes a derivative with respect to time $\tau$.
The equations of motion for the scale factors are
\begin{align}\label{eom}
\frac{d^2}{d\tau^2}\ln{a^2}=a^4-(b^2-c^2)^2
\end{align}
and permutations thereof cyclic in $a$, $b$, $c$ .
Furthermore, the  Hamiltonian constraint, $H=T+V=0$, follows from the $(0,0)$-component of the Einstein equations in vacuum.\\ 

In the case of free motion, when the potential and hence the terms on the right-hand side of~\eqref{eom} vanish, the equations of motion are easily integrated to yield:
\begin{align}\label{eq:freesol}
\beta^a=v^a\tau+\beta_0^a\,,
\end{align}
where the $v^a$ is the velocity. For the lapse function we therefore find
\begin{align}\label{abc}
N=abc\propto \exp(-\tau\sum_av^a)\,.
\end{align}
Integrating the relation between the proper time $t$ and the time $\tau$ from above and using expression~\eqref{abc} for $abc$ one obtains
\begin{align}\label{timerel}
\tau=-\frac{1}{\sum_av^a}\ln(t)+\text{const.}\,.
\end{align}
From the last two relations it then also follows that $abc\propto t$.
At this point we fix as convention that as $t\rightarrow0^+$, the singularity is approached. The same shall also be true for $\tau\rightarrow+\infty$, and hence we impose the condition $\sum_av^a>0$.

From definition~\eqref{scalefacts} of the scale factors and relation~\eqref{timerel} one easily obtains the well-known Kasner form of the free solution~\eqref{eq:freesol}:
\begin{align}
a^2\sim t^{2p_1}\,,\quad  b^2\sim t^{2p_2}\,,\quad c^2\sim t^{2p_3}\,,
\end{align}
with the relation between the Kasner exponents, $p_a$, and the velocities, $v^a$, given by:
\begin{align}\label{pdefn}
p_a=\frac{v^a}{\sum_av^a}\,.
\end{align}
The Kasner exponents $p_a$ satisfy the two further constraints
\begin{align}\label{KasnerConds}
p_1+p_2+p_3=1=p_1^2+p_2^2+p_3^2\,,
\end{align}
where the first constraint is a direct consequence of definition~\eqref{pdefn} and the second follows from the Hamiltonian constraint which is discussed below.

The kinetic term of the system in equation~\eqref{Lagr_Mixmaster} is in fact the kinetic term of a fictitious, massless particle moving in a three-dimensional $\beta$-space. A line element, $d\sigma^2$, in this space, also known as Wheeler-DeWitt minisuperspace, is given by:
\begin{align}
d\sigma^2=\sum_a(d\beta^a)^2-\Big(\sum_a d\beta^a\Big)^2\equiv G_{ab}d\beta^ad\beta^b\,,
\end{align}
which defines the metric $G_{ab}$:
\begin{align}\label{eq:supermetric}
(G_{ab})=
\left(
\begin{array}{ccc}
0 & -1 & -1\\
-1 & 0 & -1\\
-1 & -1 & 0
\end{array}
\right)\,.
\end{align}
Through a linear transformation this metric can be shown to have Lorentzian signature. 
The fictitious particle satisfies the Hamiltonian energy condition, $H=T+V=0$, which takes the form
\begin{align}
G_{ab}\dot\beta^a\dot\beta^b+V(\beta)=0\,.
\end{align}
In the case of free motion, c.f. equation~\eqref{eq:freesol}, this constraint reduces to
\begin{align}\label{nullcond}
G_{ab}v^a v^b=0\,.
\end{align}
We shall also call this the~\textit{null condition} which states that the fictitious particle is massless, moving along light-like paths in $\beta$-space in the absence of a potential.\\

It has been shown with some level of rigour that in the limit $\tau\rightarrow+\infty$, i.e. in the limit towards the singularity, the potential $V(\beta)$ written in equation~\eqref{eq:potential}, assumes a particularly simple structure and is given by a set of infinitely sharp potential walls. Here we will only state the result of this simplification and refer the reader to~\cite{DHN:reviewBKL} for details of the derivation:

The first step of simplification is that in the vicinity of the singularity the only relevant potential terms in~\eqref{eq:potential} are $a^4$, $b^4$ and $c^4$, while the other three mixed terms vanish with the potential taking the effective form $\sim e^{-4\beta^1}-e^{-4\beta^2}-e^{-4\beta^3}$. 

In the second simplification step it can be shown, by introducing hyperbolic polar coordinates, that each of the remaining potential terms of exponential form becomes an infinitely sharp wall. In order to describe this it is convenient to introduce the notion of a~\textit{wall form}, 
\begin{align}
w^{(A)}(\beta):=w^{(A)}_{\phantom{(A)}a}\,\beta^a\,,
\end{align}
where the index $A=1,2,3$ labels the three different potential walls with
$w^{(1)}=(1,0,0)$, $w^{(2)}=(0,1,0)$ and $w^{(3)}=(0,0,1)$.
The potential in the vicinity of the singularity then takes the form
\begin{align}
\lim_{\tau\rightarrow +\infty} V\sim\sum_{A=1}^3 \Theta_\infty\Big(-w^{(A)}(\beta)\Big)\,,
\end{align}
where an infinite step function is defined as:
\begin{align}\label{eq:pot_step}
\Theta_\infty(x)=\begin{cases} 
0&\text{if }\,x<0\\ 
+\infty&\text{if }\,x>0 
\end{cases}\,.
\end{align}
There is thus a sharp wall situated at the plane defined by $w^{(1)}(\beta)=0$ and similarly also for the other two walls. The fictitious, massless particle moves on light-like trajectories in the volume bounded by the sharp walls, such that at any time, $w^{(A)}(\beta)\geq0$, for each of the three walls. Away from the any walls the particle moves freely in a Kasner flight, as described by~\eqref{eq:freesol}. When the particle encounters one of the sharp potential walls it is reflected by the wall and $v^a$ is transformed into the velocity vector $v'^a$. The reflection obeys the law of geometric reflection on the hyperplane, $w^{(A)}(\beta)=0$, given by:
\begin{align}
v'^a =v^a-2\,\frac{v^b\,w^{(A)}_{\phantom{(A)}b}}{\,w^{(A)} \boldsymbol \cdot w^{(A)}\,}\,w^{(A)\,a}\,.
\end{align}
Indices are raised by the inverse of the supermetric~\eqref{eq:supermetric} and the product in the denominator is also defined with respect to that metric. The three sharp potential walls form a `billiard' table in $\beta$-space, constraining the motion of the particle. This system, in which the motion of the fictitious particle is chaotic due to the region defined by the potential walls having finite volume, has been termed~\textit{cosmological billiard}. The notion of the cosmological billiard extends to theories in higher-dimensions and with additional degrees of freedom. Furthermore, the particular arrangement of the potential walls, leads to a remarkable and unexpected connection of the cosmological billiard with infinite-dimensional Kac--Moody Lie algebras~\cite{DH_E10,DHJN_hyperbolic,DHN_E10}. While this is possibly one of the most significant further developments of the BKL conjecture, it is not of importance for the purpose of this article. 
The homogeneous, diagonal Bianchi IX model is the prototypical example for a cosmology with BKL oscillations. One may wonder however whether a similar model might also exist in $2+1$ dimensions.

\section{A toy model in $2+1$ dimensions}\label{sec:toymodel}
Following an intuitive approach, we consider a model derived from the homogeneous, diagonal Bianchi IX model by simply dropping one of the scale factors. Without loss of generality we let this be the scale factor $c$ and the resulting model is of the form
\begin{align}\label{eq:L_toymodel}
L&=-2\dot\beta^1\dot\beta^2-\frac12\left(e^{-4\beta^1}+e^{-4\beta^2}-2e^{-2(\beta^1+\beta^2)}\right)\,.
\end{align}
The associated equations of motion read:
\begin{align}\label{eq:eoms}
\begin{split}
\frac{d^2}{d\tau^2}\ln a^2&=2\left(b^4-a^2b^2\right)\,,\\
\frac{d^2}{d\tau^2}\ln b^2&=2\left(a^4-a^2b^2\right)
\end{split}
\end{align}
and the Hamiltonian constraint takes the form
\begin{align}\label{eq:HamiltonainConstraint}
(\ln a^2)^{\bm\cdot}(\ln b^2)^{\bm\cdot}&=\left(a^2-b^2\right)^2\,.
\end{align}
Recalling the discussion of $\beta$-space in the previous section, the supermetric now takes the form
\begin{align}
(G_{ab})=
\left(
\begin{array}{cc}
0 & -1 \\
-1 & 0 
\end{array}
\right)\,,
\end{align}
with indices $a,b=1,2$ and which also has Lorentzian signature. In the case of free motion, where the right-hand side of the equations of motion are zero, we again obtain the standard Kasner solution, $\beta^a=v^a\tau+\beta^a_0$. From the null constraint
\begin{align}\label{eq:nullconstraint}
G_{ab}v^av^b=-2v^1v^2=0
\end{align}
we see that one of the components of the velocity vector $v^a$ are necessarily zero.

We will now rewrite the equations of motion~\eqref{eq:eoms} in a few steps to obtain a form where we have first order time-derivatives on the left-hand side. This allows us to draw a phase portrait of the system from which qualitative information about the system's dynamics can be extracted. A more detailed version of the following derivation is given in appendix~\ref{app:dynamical_system}.
\subsection{Dynamics in phase space}\label{sec:phasespace} 

\begin{figure}[p!]
    \centering
    \begin{subfigure}[t]{\textwidth}
    \centering
        \includegraphics[width=0.8\linewidth]{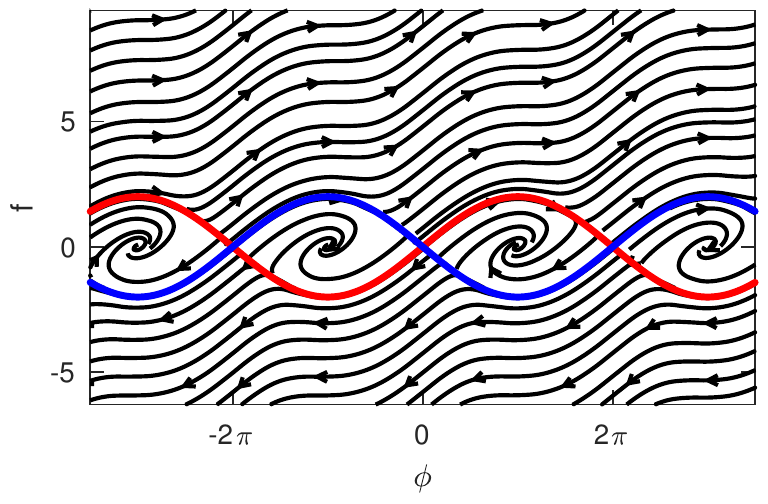} 
        \caption{Phase portrait of the dynamical system~\eqref{eq:vector_field} with a few integral curves shown explicitly. Arrows point in direction of decreasing $\eta$ (movement away from the singularity). The two separatrix curves~\eqref{eq:portrait_equation} are shown in red and blue.} \label{fig:dynplots}
    \end{subfigure}
    
    \vspace{1cm}
        \begin{subfigure}[t]{0.45\textwidth}
        \centering
        \includegraphics[width=\linewidth]{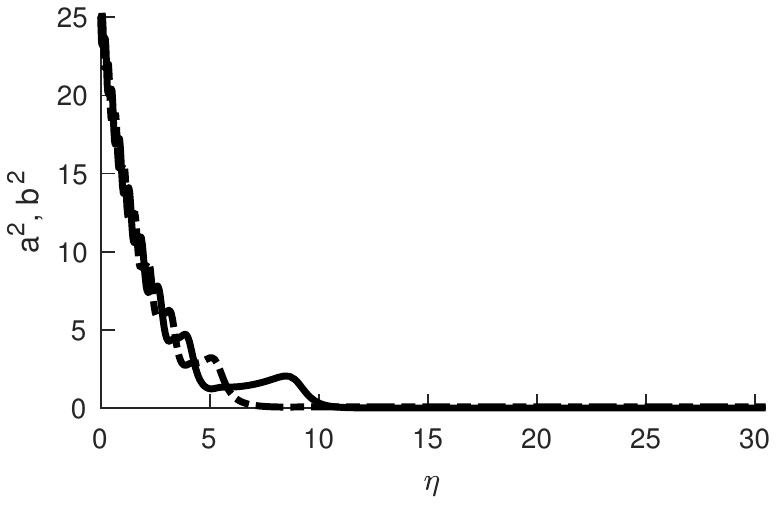} 
       \caption{Time evolution of the scale factors $a^2$ (solid) and $b^2$ (dashed).} \label{fig:a2b2}
    \end{subfigure}
    \hfill
    \begin{subfigure}[t]{0.45\textwidth}
        \centering
        \begin{tikzpicture}
   \node[anchor=south west,inner sep=0] (image) at (0,0) {\includegraphics[angle=45,width=1.\textwidth]{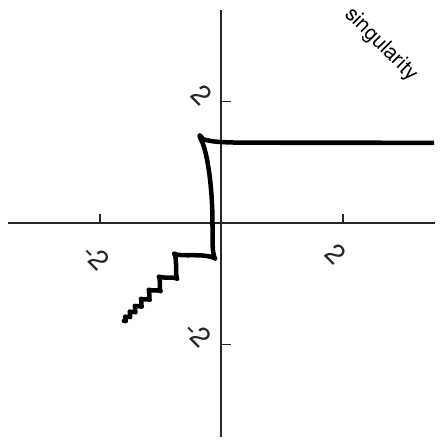}};
    \begin{scope}[x={(image.south east)},y={(image.north west)}]
    	\node[below] at (0.85,0.85) {\small$\beta^1$};
	\node[below] at (0.15,0.85) {\small$\beta^2$};
	\draw[decoration = {zigzag,segment length = 1.5mm, amplitude = .3mm},decorate] (0.28,0.8385)--(0.7,0.8385);
    \end{scope}
\end{tikzpicture}
        \caption{Trajectory in $\beta$-space: particle traveling from bottom to top, towards the singularity located at infinity inside the upper quadrant.}
                \label{fig:beta_dynamics}
    \end{subfigure}
        \caption{Toy model dynamics illustrated in different representations.}   
\end{figure}
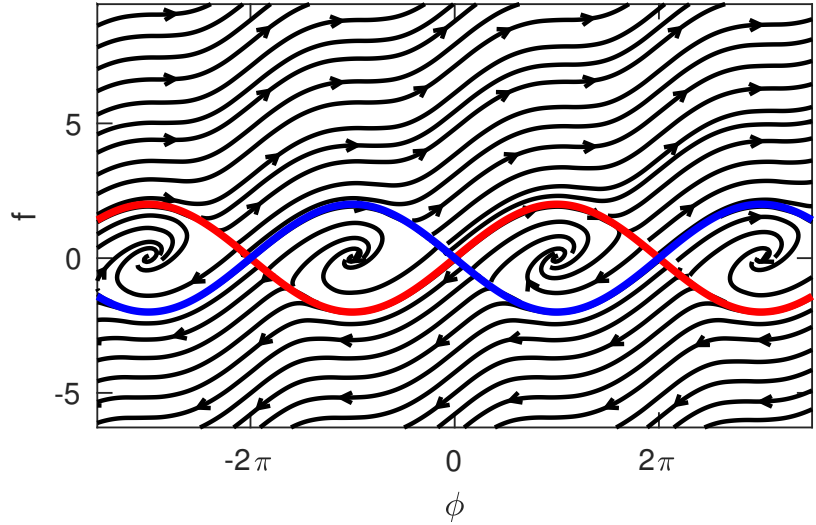
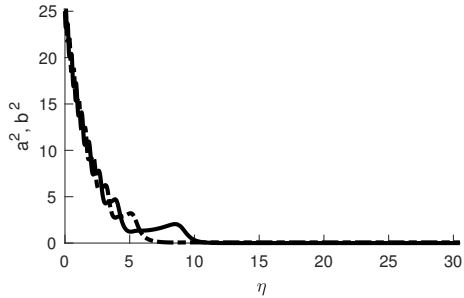
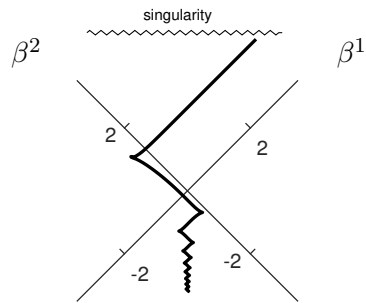

We start by defining new, time-dependent variables $u$, $v$, $z$ and $w$:
\begin{align}
a^2=\frac{z+w}{2}\,,\quad b^2=\frac{z-w}{2}\,,\nn
\end{align}
and
\begin{align}
\left(\log a^2\right)^{\bm \cdot}=u+v\,,\quad \left(\log b^2\right)^{\bm \cdot}&=u-v\,,\nn
\end{align}
from which the following set of equations is obtained:
\begin{alignat*}{3}
\dot z &=uz+vw\,,\quad&&\dot w &&=uw+vz\,,\nn\\
\dot v &=-zw\,, &&\dot u&&=w^2\,.
\end{alignat*}
The Hamiltonian constraint takes the form $u^2-v^2=w^2$. In a second step of redefinitions we introduce the functions $f$, $R$ and $\phi$ as variables:
\begin{alignat*}{3}
z&=\frac{f\cdot R}{2}\,,\quad &&w&&=R\,\cos\left(\frac{\pi-\phi}{2}\right)\,,\nn\\
u&=R\,, &&v&&=R\,\sin\left(\frac{\pi-\phi}{2}\right)\nn
\end{alignat*}
and we also introduce a new time variables $\eta$ such that $d\eta/d\tau=R$. Then we obtain, after a few steps of algebra, the following equations:
\begin{align}\label{eq:vector_field}
\begin{split}
\phi'&=f+\sin(\phi)\,,\\
f'&=\frac f2\Big(1+\cos(\phi)\Big)+\sin(\phi)\,,
\end{split}
\end{align}
where the prime denotes a derivative with respect to the time $\eta$. Eliminating time from these two equations we obtain
\begin{align}\label{eq:portrait_equation}
\frac{df}{d\phi}=\frac{q(1+\cos(\phi))+2\sin(\phi)}{2(q+\sin(\phi))}\,.
\end{align}
The phase portrait of this system is shown in Figure~\ref{fig:dynplots}. The arrows on the integral curves point in the direction of increasing proper time $t\rightarrow+\infty$, i.e. $\tau\rightarrow-\infty$.

There are two types of fixed points in this system, see appendix~\ref{sec:lin_stab}. The first type are spiral sinks located at $(\phi,f)=((2n+1)\pi,0)$, where $n$ is integer. The second type, located at  $(2\pi n,0)$, are non-hyperbolic fixed points. Analysis shows that these fixed points are topologically equivalent to standard saddle points.  Furthermore, these fixed points are connected by separatrix curves and it is easy to check, by substitution into equation~\eqref{eq:portrait_equation}, that the analytic expressions for the separatrix curves are
\begin{align}\label{eq:separatrix}
f=\pm2\sin\left(\frac\phi2\right)\,.
\end{align}
We are then left with the following picture: for $\tau\rightarrow+\infty$, integral curves will approach the separatrix and will follow its periodic form, either above or below the region enclosed by the red and blue curves in Figure~\ref{fig:dynplots}, indefinitely. This behaviour represents the analogue of the BKL oscillatory regime in this toy model system.
Figure~\ref{fig:a2b2} shows the evolution of the scale factors $a^2$ and $b^2$ as we start at a point far from the separatrix and follow its approach, again for $\tau\rightarrow+\infty$. Note that the BKL oscillatory regime is not resolved by the numerics.

%BEGIN FIGURE: PHASE SPACE, A2B2 AND BETA SPACE FOR THE TOY MODEL
% END FIGURE

\subsection{Dynamics in $\beta$-space}\label{sec:betaspace}
Up to now we have introduced a toy model and represented its dynamics in phase space. Now we focus on how this dynamics looks like in the space spanned by the two exponents $\beta^1$ and $\beta^2$, which the scale factors $a$ and $b$ are functions of, respectively. We discuss the trajectory of a fictitious, massless particle moving in this $\beta$-space with coordinates $(\beta^1,\beta^2)$.

In Figure~\ref{fig:beta_dynamics} we show an exemplary trajectory of the particle in $\beta$-space. The singularity is located in the upper quadrant, at infinity. The trajectory of the particle shown, corresponds to an integral curve in the phase portrait~\ref{fig:dynplots} starting at a large value of $f$ and then approaches the separatrix curve. 
In Figure~\ref{fig:beta_dynamics} the particle travels from bottom to top towards the singularity as time $\tau$ increases. In the lower quadrant, the particle is undergoing a bouncing motion.
Each time the particle crosses the line $\beta^1=\beta^2$, where the potential is zero, its trajectory is light-like, as required by the null constraint~\eqref{eq:nullconstraint}.
We deduce from the Hamiltonian constraint with non-zero value of the potential term, that away from this line the particle's trajectory becomes time-like.
As the particle approaches the singularity, the points along the trajectory at which bounces occur move further and further to the left and right from the line, $\beta^1=\beta^2$. We emphasise that this bouncing motion is present away from the singularity. 

As we have seen in the discussion of the toy model's phase space, the oscillatory dynamics continues to persist in the vicinity of the singularity when integral curves closely follow the separatrix. We have also discussed in section~\ref{sec:BIXreview} for the Bianchi IX model that in this limit the behaviour of the particle's velocity vector is governed by a geometric reflection law. We note that in the case of our toy model, the correct transformation of the velocity vector may similarly be realised in terms of such a reflection law, when the potential walls are time-like. Such a potential wall is represented by a time-like hyperplane in $\beta$-space orthogonal to the space-like vector $w_a=(1,-1)$. The effect of a time-like wall on the particle's velocity vector upon collision is well-known, see for example~\cite{DHN:reviewBKL}. More precisely, by decomposing the particle's velocity vector into a component parallel and a component orthogonal to the wall it is not hard to see that upon collision the parallel component stays the same, while the orthogonal component has its sign reversed. As a result, starting in our case with velocity vector $(v,0)$, the velocity after the collision will be $(0,v)$. As in the case of the Bianchi IX model, the collision is then governed by the law of geometric reflection on the hyperplane defined by $w(\beta)=0$ which we repeat here for the reader's convenience
\begin{align}
v'^a =v^a-2\frac{v^bw_b}{w\boldsymbol \cdot w}w^a\,.
\end{align}
We may also represent this behaviour in terms of Kasner exponents defined as $p_a=v^a/(v^1+v^2)$, in analogy with equation~\eqref{pdefn}. The sequence of values assumed by the Kasner exponents is
\begin{align}
\left(
\begin{array}{c}
p_1 \\
p_2
\end{array}
\right)=
\left(
\begin{array}{c}
1\\
0 
\end{array}
\right)\rightarrow
\left(
\begin{array}{c}
0\\
1 
\end{array}
\right)\rightarrow
\left(
\begin{array}{c}
1 \\
0 
\end{array}
\right)\rightarrow\ldots\,.
\end{align}

\begin{figure}[!t]
        \centering
        \begin{tikzpicture}
   \node[anchor=south west,inner sep=0] (image) at (0,0) {\includegraphics[angle=0,width=0.16\textwidth]{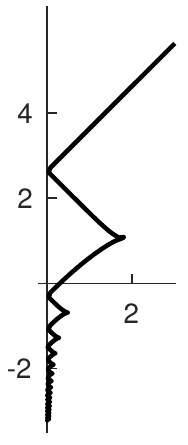}};
    \begin{scope}[x={(image.south east)},y={(image.north west)}]
    	\node[below] at (0.3,1.1) {$\beta^1+\beta^1$};
	\node[below] at (1.3,0.425) {$\beta^1-\beta^2$};
    \end{scope}
\end{tikzpicture}
        \caption{Trajectory of the particle in $\beta$-space. Using the symmetry of the toy model under exchange of $\beta^1$ and $\beta^2$ we may display the dynamics in one half of $\beta$-space.}
                \label{fig:beta_dynamics_half} 
\end{figure}

Finally, to complete our discussion of dynamics in $\beta$-space we make a simple observation which will turn out to have a physical interpretation discussed in the following next section: considering the Lagrangian~\eqref{eq:L_toymodel} of the toy model we see that it is invariant under a simple exchange of the variables $\beta^1$ and $\beta^2$. As a result the $\beta$-space possesses a reflection symmetry about the line defined by, $\beta^1=\beta^2$, which makes it convenient to display the particle's trajectory in only one half of $\beta$-space in the way shown in Figure~\ref{fig:beta_dynamics_half}.
% BEGIN FIGURE: HALFED BETA SPACE
% END FIGURE

Up to this point we have treated the toy model abstractly and even its initial conception has been intuitive. In the following section we will turn to a discussion of gravity in $2+1$ dimensions. As briefly explained in the introduction, counter to first intuition, $2+1$-dimensional gravity is not necessarily a trivial theory due to the possibility of including topological degrees of freedom. We will discuss such a scenario and propose a connection with our toy model. 
\section{Towards a physical interpretation}~\label{sec:physical_interpretation}
Over several decades the study of both classical and quantum gravity in $2+1$ dimensions has developed into a vast subject of widely varying approaches and technically different formulations, see for instance~\cite{Deser:1983tn,Witten:1988hc,Moncrief:1989dx} and~\cite{Carlip:1995zj} for a broad review. Here we shall focus on a well-known model for $2+1$-dimensional gravity, written in the Arnowitt--Deser--Misner (ADM) formalism. This allows us to make a direct link to our model with oscillatory dynamics discussed in the preceding section.

\begin{figure}
\centering
\resizebox{0.5\textwidth}{!}{
\begin{tikzpicture}[domain=-0.3:5]

%coordinates axes
\draw[->,] (-1.5,0) -- (3,0);
\draw[->,] (0,-0.1) -- (0,2); 

%parallelogram
\draw[-,] (1,0) node[below]{\tiny$1$} -- (1.5,1);
\draw[-,] (1.5,1) node[above]{\tiny$\rho+1$} -- (0.5,1);
\draw[-,] (0.5,1) node[above]{\tiny$\rho$} -- (0,0);

\node[] at (2.8, 1.8) {\scriptsize$\mathbb H$};

\end{tikzpicture}
}   
\caption{Representation of the torus in the upper half-plane. Opposite sides of the parallelogram are identified.}
\label{fig:parallelogram}
\end{figure}
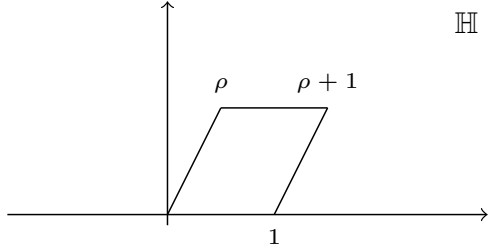

\begin{figure}
\centering
\resizebox{0.6\textwidth}{!}{
\begin{tikzpicture}[domain=-0.3:5]

%coordinates axes
\draw[->,] (-2,0) -- (2,0) node[right]{\tiny$\nu$};
\draw[->,] (0,-0.2) -- (0,2.8) node[above]{\tiny$e^\varphi$};

%fundamental domain
\draw[-,thick] (-1/2,0.8660254) -- (-1/2,2.4);
\node[left] at (-1/2, 2.4) {\tiny $\mathcal F$};
\draw[-,thick] (1/2,0.8660254) -- (1/2,2.4);
\draw [black,dashed,thin,domain=0:60] plot ({cos(\x)}, {sin(\x)});
\draw [black,thick,domain=60:120] plot ({cos(\x)}, {sin(\x)});
\draw [black,dashed,thin,domain=120:180] plot ({cos(\x)}, {sin(\x)});

\node[below] at (1, 0) {\tiny $1$};
\node[below] at (-1, 0) {\tiny $-1$};
\node[label={[label distance=0.1cm]235:\tiny$i$}] at (0.25, 1.25) {};

\draw[<->,very thick] (0,1) -- (0,2);
\node[left] at (0, 2.) {\tiny$Q$};
\tkzDefPoint(0,1.4){Q}
\tkzLabelPoint[left](Q){}
\foreach \n in {Q}
  \node at (\n)[circle,fill,inner sep=1.1pt]{};

\end{tikzpicture}
}   
\caption{Motion of the particle in moduli space. The particle bounces up-and-down along the imaginary axis inside the fundamental domain, between the point at $i$ and a point $Q$, while $Q$ is moving upwards as the singularity is approached.}
\label{fig:moduli_space_dynamics}
\end{figure}
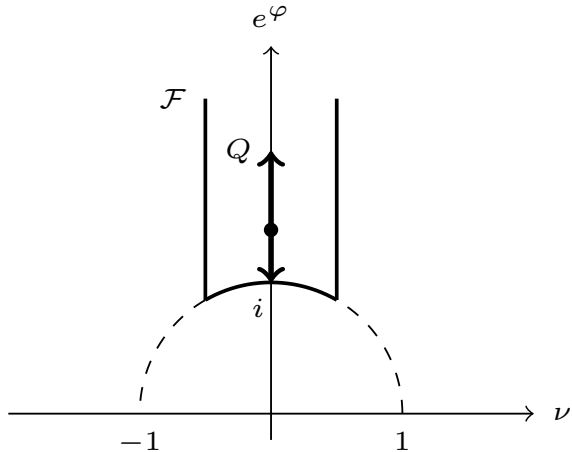

The model that we discuss is the $2+1$-dimensional torus universe where the space-time topology is, $\mathbb R\times T^2$. In different forms this case has for instance been treated in~\cite{Moncrief:1989dx,Hosoya:1989yj,Waldron_torus_milne}.
The first step is to find a suitable parameterisation of the metric on the torus topology. It is a simple fact that a torus can be represented by a parallelogram in the upper half-plane $\mathbb H$, as shown in Figure~\ref{fig:parallelogram}, with opposite sides of the parallelogram identified. To obtain the moduli space of the torus one must account for global diffeomorphisms. These are transformations of the complex (Teichm\"uller) parameter, $\rho=\nu+ie^{\varphi}$, which lead to equivalent tori. The group of these transformations is the modular group $PSL(2,\mathbb Z)$ which acts on the complex parameter $\rho$ according to
\begin{align}
\left(
\begin{array}{cc}
a & b\\
c & d
\end{array}
\right)\bm\cdot\rho\rightarrow\frac{a\rho+b}{c\rho+d}\;,
\end{align}
where the matrix is an element of $PSL(2,\mathbb Z)$, i.e. $a,b,c,d\in\mathbb{Z}$ and $ad-bc=1$ and matrices differing only by a sign are considered identical.
The actions of the two generating transformation of the modular group result in $\rho\rightarrow-1/\rho$ and $\rho\rightarrow\rho+1$. The moduli space of the torus is obtained as the quotient of the upper half-plane and the modular group. It is given by the fundamental domain $\mathcal F$, shown in Figure~\ref{fig:moduli_space_dynamics}, with any point in the upper half-plane being related to a point in the fundamental domain via a modular transformation. 
%BEGIN FIGURE: PARALLELOGRAM IN THE UHP
% END FIGURE

As in section~\ref{sec:BIXreview} we decompose space-time according to
\begin{align}
ds^2=-(Ndx^0)^2 + g_{ij}dx^idx^j\,,
\end{align}
where now $i,j=1,2$.
The metric on a torus topology can be parameterised as:
\begin{align}\label{eq:spatial_metric_moduli}
g_{ij}=e^{-\lambda} \bar g_{ij}(\rho)\,,
\end{align}
where $\bar g_{ij}$ is a function of the complex modulus, $\rho$, of the torus. Furthermore, the metric $\bar g_{ij}$ is defined to have unit determinant, $\det \bar g=1$, and a volume factor $\exp(-\lambda)$.

It is a well-known fact that the space of metrics on a torus which are conformally flat, is parameterised by the $SL(2,\mathbb R)/SO(2)$ coset. We will then parameterise the spatial metric by
\begin{align}
\bar g=\mathcal{V}\mathcal{V}^T=
\left(
\begin{array}{cc}
e^{\varphi}+\nu^2e^{-\varphi} & \nu e^{-\varphi}  \\
\nu e^{-\varphi}  & e^{-\varphi} \\
\end{array}
\right)\,,
\end{align}
such that, $\det \bar g=1$, and the Vielbein, $\mathcal{V}$, is of Iwasawa decomposed form: 
\begin{align}
\mathcal{V}=
\left(
\begin{array}{cc}
1 &  \nu \\
 0 & 1 \\
\end{array}
\right)
\left(
\begin{array}{cc}
e^{\varphi /2} & 0 \\
0 & e^{-\varphi /2} \\
\end{array}
\right)\,.
\end{align}

The ADM action~\cite{ADM} written for $2+1$ dimensions with pseudo-Gaussian gauge choice, takes the form:
\begin{align}\label{eq:metricLagr}
S=\int d x^0\,d^2xN\sqrt{g}\left(K^{ij}K_{ij}-K^2-R(g)\right)\,,
\end{align} 
where $R(g)$ is the Ricci scalar of the spatial metric~\eqref{eq:spatial_metric_moduli} and $K$ is the trace of the second fundamental form $K_{ij}$. In our case $K_{ij}$ is given by
\begin{align}
K_{ij}=-\frac{1}{2N} g_{ij\,,x_0}\,.
\end{align}
With this relation, the action takes the form
\begin{align}
S=\int d x^0\,d^2x\left(\frac{\sqrt{g}}{4N}\left[g^{ik}g^{jl}-g^{ij}g^{kl}\right]g_{ij\,,x_0}g_{kl\,,x_0}-N\sqrt{g}R(g)\right)\,.
\end{align}
Fixing the lapse function to our previous choice of $N=\sqrt{g}$, we then obtain:
\begin{align}\label{eq:action}
S=\int d \tau\,d^2x\left(\frac{1}{4}\left[g^{ik}g^{jl}-g^{ij}g^{kl}\right]\dot g_{ij}\dot g_{kl}-gR(g)\right)\,.
\end{align}
The first term in the integrand is the kinetic term and the second term $gR$ represents a potential.
Evaluating equation~\eqref{eq:action} for the metric $g(\lambda,\rho)$, then yields a kinetic term of the form:
\begin{align}\label{eq:moduli_kinetic_term}
T_\mathcal{M}=\frac12\left(- \dot\lambda^2 + e^{-2\,\varphi}\dot\nu^2+\dot\varphi^2\right)\,,
\end{align}
where $\mathcal M$ stands for moduli.
The second and third term on the right-hand side form the kinetic term of a fictitious particle moving in the Poincar\'e upper half-plane.\footnote{Some readers may be more familiar with the form of this kinetic term and the associated Poincar\'e metric when written in terms of the variables $\rho_1:=\nu$ and $\rho_2:=\exp(\varphi)$.}

The evaluation of the potential term in terms of $\nu$ and $\varphi$, the $SL(2,\mathbb R)/SO(2)$ coset variables, and $\lambda$ yields a somewhat lengthy expression which we provide in appendix~\ref{app:curvature_potential}. For our purpose it is enough to state an abbreviated form of this expression\footnote{The general form of the curvature potential in space-time dimensions greater than three has been discussed in section 6.2 of~\cite{DHN:reviewBKL}. However, although we have not verified this in detail, their derivation appears to also hold for the present case of three space-time dimensions. The general structure of the potential obtained in this way appears to agree with the one we find. A detailed comparison with our expression is however not possible, since~\cite{DHN:reviewBKL} do not state the complicated coefficients explicitly.}:
\begin{align}\label{eq:ADM_potential}
V_\mathcal{M}=gR=-e^{-\lambda}&\Big(
\;\,\left[\partial_x\varphi^2+\partial_x\lambda\partial_x\varphi-\partial^2_{xx}\varphi-\partial^2_{xx}\lambda\right]\,e^{-\varphi}\nn\\
&+\left[\partial_y\varphi^2-\partial_y\lambda\partial_y\varphi+\partial^2_{yy}\varphi-\partial^2_{yy}\lambda\right]\,e^{\varphi}\nn\\
&+\left[\text{sum of coefficients involving } \nu\text{, }\partial \nu\text{, }\partial^2\nu\right]\,e^{-\varphi}
\Big)\,.
\end{align}
Having computed the Lagrange function, $L_\mathcal M=T_\mathcal M-V_\mathcal M$, for the torus moduli, we will now propose a connection with cosmological models that display oscillatory BKL dynamics.
%BEGIN FIGURE: MODULI SPACE
% END FIGURE

\subsection{Dynamics in moduli space}
We start with a discussion of the kinetic terms. For the purpose of this article we will restrict our discussion to the case of a diagonal spatial metric, i.e. when $\nu=0$ and hence also $\dot\nu=0$. Below we will further remark on this restriction. Then, considering the kinetic term for the torus moduli~\eqref{eq:moduli_kinetic_term}, we note that by making the change of variables
\begin{align}\label{eq:lambda_varphi}
\begin{split}
\lambda&=2(\beta^1+\beta^2)\,,\\
\varphi&=2(\beta^1-\beta^2)\,,
\end{split}
\end{align}
we recover, up to a constant factor of one half, the kinetic term, $T=-2\dot\beta^1\dot\beta^2$, of our toy model example~\eqref{eq:L_toymodel}, i.e. $T_\mathcal{M}=2T$.
This is the basic connection between the minisuperspace degrees of freedom and the topological degrees of freedom of gravity in $2+1$ dimensions that we propose. 

At the end of section~\ref{sec:betaspace} we pointed out the invariance of the toy model Lagrangian~\eqref{eq:L_toymodel} under exchange of $\beta^1$ and $\beta^2$, implying the existence of a potential wall at $\beta^1=\beta^2$, see Figure~\ref{fig:beta_dynamics_half}. There is a direct correspondence of this symmetry in moduli space. Namely, we see from a simple computation, that the effect of the generating modular transformation, $\rho'=-1/\rho$, on the variable $\varphi=\ln\left(\,\text{Im}(\rho)\,\right)$, is simply to reverse its sign:
\begin{align}\label{eq:varphi_transf}
\varphi'=\ln\left(\,\text{Im}(\rho')\,\right)=\ln\left(\frac{e^\varphi}{\nu^2+e^{2\varphi}}\right)=\ln\left(e^{-\varphi}\right)=-\varphi\,.
\end{align} 
In the second equality we have kept $\nu$ explicitly only for clarity of explanation (recall, $\nu=0$). By identification of $\varphi$ with the mini-superspace degrees of freedom in equation~\eqref{eq:lambda_varphi}, the change in sign is equivalent to an exchange of $\beta^1$ and $\beta^2$.
This shows that the reflection symmetry present in $\beta$-space is directly related to the invariance of the torus under this modular transformation which is not broken in the toy model.
 
While we will discuss the potential term, $V_\mathcal M$, in a moment, it is interesting to consider the trajectory of a fictitious particle in moduli space, governed by our toy model. With the above changes of variables the toy model~\eqref{eq:L_toymodel} takes the form:
\begin{align}\label{eq:L_toymodel_moduli}
2\,L&=\frac12\left(-\dot\lambda^2+\dot\varphi^2\right)-2\left(e^{-(\lambda+\varphi)}+e^{-(\lambda-\varphi)}-2e^{-\lambda}\right)\,.
\end{align}
The particle's trajectory is confined to the imaginary axis, $i\exp(\varphi)$ with $\varphi\geq0$, where it performs a bouncing motion, see Figure~\ref{fig:moduli_space_dynamics}.
Each time the particle bounces on the line $\beta^1-\beta^2=0$ line in Figure~\ref{fig:beta_dynamics_half}, the particle in moduli space reaches the point $i$ in the upper half-plane and bounces back up. The following bounce, directing the particle back down, occurs at a point $Q$ along the imaginary axis which moves upwards to infinity as the particle approaches the singularity.

Next, we turn our attention to the potential terms. Consider the general potential term~\eqref{eq:ADM_potential} derived in ADM formalism.
We note that the potential of our toy model does~\textit{not} fit this form, since a term with only a $\exp(-\lambda)$-factor is not present. There was of course no reason to believe that the toy model potential should match the form of the potential derived from the ADM action. However, we note that the $\exp(-\lambda)$-term has no significant impact on the existence of oscillatory dynamics. Due to its negative coefficient, its effect is to repel the particle form the singularity, while oscillations are solely due to the other two terms with $\varphi$ in the exponential.
We postpone an exhaustive discussion of the potential~\eqref{eq:ADM_potential} with its complicated coefficient functions to another publication. Let us finally also remark on the restriction to the case of $\nu=0$ and $\dot\nu=0$. While this restriction might seem arbitrary, at least the restriction to $\dot\nu=0$ is not without reason. Namely, in the context of cosmological billiards it has been shown in~\cite{DHN:reviewBKL}, that in the limit towards the singularity all off-diagonal degrees of freedom in the Iwasawa decomposition of the metric will `freeze' in the sense that their time derivatives will be zero. While the derivation of the phenomenon of `freezing' presented in~\cite{DHN:reviewBKL} also holds in $2+1$ dimensions, its full consequence are an issue to be investigated. In particular a non-zero value of $\nu$ will have an impact on the symmetry transformation~\eqref{eq:varphi_transf}. 
\section{Conclusion}
Our starting point has been the question whether there are cosmological models in $2+1$ space-time dimensions which display an analogue of BKL oscillations. In this preprint we have outlined the main ideas, pointing to an affirmative answer to this question. While we base most of our reasoning on a toy model, it seems likely that further investigation of cosmological models for the torus moduli will confirm our conclusions. We believe that this adds an interesting new aspect to the study of $2+1$-dimensional gravity.  

\section*{Acknowledgements}
We are grateful to Axel Kleinschmidt for comments on a draft of this paper. P. F. would like to thank Thibault Damour for discussions on gravity in $2+1$ space-time dimensions.
P. F. acknowledges support from the CARMIN Post-Doctoral Fellowship programme (funding period 2014-2016) and would like to thank the Institut des Hautes \'Etudes Scientifiques and the Max Planck Institute for Gravitational Physics for hospitality during the initial stages of this work.
\appendix
\section{Dynamical system analysis}\label{app:dynamical_system} 
First we write the equations of motion~\eqref{eq:eoms} in terms of the variables $z$, $w$, $u$ and $v$ by making the following re-definitions
\begin{align}
a^2&=\frac{z+w}{2}\,,\quad
b^2=\frac{z-w}{2}
\end{align}
and
\begin{align}
\left(\log a^2\right)_\tau=u+v\,\quad \left(\log b^2\right)_\tau&=u-v\,.
\end{align}
In terms of the variables, the equations of motion read
\begin{align}
\begin{split}
u_\tau+v_\tau&=(w-z)w\,,\\
u_\tau-v_\tau&=(w+z)w\,.
\end{split}
\end{align}
and we also find the two further relations
\begin{align}
\begin{split}
z_\tau-w_\tau&=zu-zv-uw+wv\,,\\
z_\tau+w_\tau&=zu+zv+uw+wv\,.
\end{split}
\end{align}
From these we deduce the following set of equations
\begin{align}\label{sys1}
\begin{split}
u_\tau&=w^2\,,\\
v_\tau&=-zw\,,\\
z_\tau&=zu+vw\,,\\
w_\tau&=zv+uw\,,
\end{split}
\end{align}
together with the Hamiltonian constraint~\eqref{eq:HamiltonainConstraint} given by $u^2-v^2=w^2$.
We proceed to write~\eqref{sys1} in terms of the variables $\gamma$, $F$ and $R$, by making the transformations
\begin{alignat*}{2}
u&=R\,, \quad &&v=R\sin(\gamma)\,,\\
z&=FR\,, \quad &&w=R\cos(\gamma)\,.
\end{alignat*}
The system~\eqref{sys1} then takes the form
\begin{align}\label{sys2}
R_\tau&=R^2\cos^2(\gamma)\,,\nn\\
R_\tau\sin(\gamma) + R\gamma_\tau\cos(\gamma)&=-FR^2\cos(\gamma)\,,\nn\\
F_\tau R+FR_\tau&=FR^2+R^2\sin(\gamma)\cos(\gamma)\,,\nn\\
R_\tau\cos(\gamma)-R\gamma_\tau\sin(\gamma)&=FR^2\sin(\gamma)+R^2\cos(\gamma)\,.
\end{align}
By re-parameterising the intermediate time, $\tau$, according to
\begin{align}
d\tau=\frac{d\eta}{R}\,,
\end{align}
we make the equations linear in $R$ and furthermore re-write the system in the following way
\begin{align}\label{sys2}
\begin{split}
R_\eta&=R\cos^2(\gamma)\,,\\
F_\eta&=F\sin^2(\gamma)+\sin(\gamma)\cos(\gamma)\,,\\
\gamma_\eta&=-F-\sin(\gamma)\cos(\gamma)\,,\\
\frac{d\eta}{d\tau}&=R\,.
\end{split}
\end{align}
Finally we give expressions for $a^2$, $b^2$ in terms of the new variables
\begin{align}
a^2&=\frac{R\left(F+\cos(\gamma)\right)}{2}\,,\quad b^2=\frac{R\left(F-\cos(\gamma)\right)}{2}\,,
\end{align}
We now show that for the dynamical system~\eqref{sys2} there exists a separatrix solution and we will give its form explicitly.

Defining $f:=F/2$ and $\delta:=2\gamma$ and making the shift, $\phi:=\pi-\delta$, the second and third equation of~\eqref{sys2} take the form
\begin{align}\label{sys3}
\begin{split}
f_\eta&=\frac{f}{2}\left(1+\cos(\phi)\right)+\sin(\phi)\,,\\
\phi_\eta&=f+\sin(\phi)\,.
\end{split}
\end{align}
or, after eliminating the time $\eta$, we obtain
\begin{align}
\frac{df}{d\phi}=\frac{f(1+\cos(\phi))+2\sin(\phi)}{2(f+\sin(\phi))}\,.
\end{align}
An exact solution of this equation is given by
\begin{align}
f=\pm 2\sin\left(\frac{\phi}{2}\right)\,.
\end{align}
This is the separatrix solution which is the boundary, separating two different modes of behaviour of our system.
\subsection{Linear stability analysis}\label{sec:lin_stab}
The system linearised around the point $(\phi_0,f_0)$ takes the form
\begin{align}
\left(
\begin{array}{c}
\phi' \\
f'
\end{array}
\right)
=\left(
\begin{array}{cc}
 1 & \cos(\phi_0)\\
 \frac{1+\cos(\phi_0)}{2} & \cos(\phi_0)
\end{array}
\right)
\left(
\begin{array}{c}
 \phi \\
 f
\end{array}
\right)
\end{align}
The for the two types of fixed points we find:
\begin{itemize}
\item $(\phi_0,f_0)=((2n+1)\,\pi,0)$: the system has two complex eigenvalues
\begin{align}
\lambda_1=\frac{-1+i\sqrt 3}{2}\;\text{  and  }\;\lambda_2=\frac{-1-i\sqrt 3}{2}
\end{align}
with negative real parts. The fixed points are thus spiral sinks.
\item $(\phi_0,f_0)=(2\pi\,n,0)$: the system has eigenvalues
\begin{align}
\lambda_1=0\;\text{  and  }\;\lambda_2=2\,.
\end{align}
and corresponding eigenvectors:
\begin{align}
\vec v_1=\frac{1}{\sqrt 2}\left(
\begin{array}{c}
 -1 \\
 1
\end{array}
\right)\;
\text{  and  }\;\vec v_2=\frac{1}{\sqrt 2}\left(
\begin{array}{c}
 1 \\
 1
\end{array}
\right)\,.
\end{align}
Since the real part of one of the eigenvalues is zero, this fixed point is of non-hyperbolic type. Analysis shows however, that the fixed points are topologically equivalent to usual saddle points.
\end{itemize}

\section{Curvature potential}\label{app:curvature_potential}
The potential term of action~\eqref{eq:action} is given by $gR(g)$. It is easy to show that in two dimensions the Ricci scalar is given by a single component of the Riemann tensor, $R(g)=2/g\cdot R_{xyxy}$, where $x:= x^1$, $y:= x^2$. The potential, given by $gR=2R_{xyxy}$, in explicit form reads
 \begin{align}
V_\mathcal M=gR=-e^{-\lambda}&\Big(
\;\,\,\left[\partial_x\varphi^2+\partial_x\lambda\partial_x\varphi-\partial^2_{xx}\varphi-\partial^2_{xx}\lambda\right]\,e^{-\varphi}\nn\\
&+\left[\partial_y\varphi^2-\partial_y\lambda\partial_y\varphi+\partial^2_{yy}\varphi-\partial^2_{yy}\lambda\right]\,e^{\varphi}\nn\\
&+\big[ - 2\,\partial^2_{xy}\nu + \partial_{x}\lambda\,\partial_{y}\nu + \partial_{y}\lambda\,\partial_{x}\nu \nn\\
&+ 3\,\partial_{x}\varphi\,\partial_{y}\nu + \partial_{y}\varphi\,\partial_{x}\nu  + 2\,\partial^2_{xy}\lambda\,\nu \nn\\
&+ 2\,\partial^2_{xy}\varphi\,\nu + 2\,\partial^2_{yy}\nu\,\nu - \partial^2_{yy}\lambda\,\nu^2\nn\\
& - \partial^2_{yy}\varphi\,\nu^2 + 2\,\partial_{y}\nu^2 + \partial_{y}\varphi^2\,\nu^2 \nn\\
&- \partial_{x}\lambda\,\partial_{y}\varphi\,\nu - \partial_{y}\lambda\,\partial_{x}\varphi\,\nu - 2\,\partial_{y}\lambda\,\partial_{y}\nu\,\nu\nn\\
&- 2\,\partial_{x}\varphi\,\partial_{y}\varphi\,\nu - 4\,\partial_{y}\varphi\,\partial_{y}\nu\,\nu + \partial_{y}\lambda\,\partial_{y}\varphi\,\nu^2\big]\,e^{-\varphi}
\Big)\,.
\end{align}  

%%%% BIBLIOGRAPHY
{\small
\bibliography{BKL_toymodel}
\bibliographystyle{utphys}
}

\end{document}